\documentclass[11pt]{article}
\usepackage[margin=1in]{geometry}
\usepackage{graphicx}
\usepackage{subcaption}
\usepackage{booktabs}
\usepackage{hyperref}
\usepackage{amsmath}
\usepackage{siunitx}
\usepackage{caption}
\title{\textbf{Investigating the Association Between Text-Based Indications of Foodborne Illness from Yelp Reviews and New York City Health Inspection Outcomes (2023)}}
\author{
\textbf{Eden Shaveet}$^{1}$\thanks{Equal contribution.} \quad
\textbf{Crystal Su}$^{1}$\footnotemark[1] \quad
\textbf{Daniel Hsu}$^{1}$ \quad
\textbf{Luis Gravano}$^{1}$\\[4pt]
$^{1}$Columbia University, Department of Computer Science, New York, NY, USA\\
\texttt{\{ems2349, ys3791\}@columbia.edu \quad \{djhsu, gravano\}@cs.columbia.edu}
}
\date{}

\begin{document}
\maketitle

\begin{abstract}
Foodborne illnesses are gastrointestinal conditions caused by consuming contaminated food. Restaurants are critical venues to investigate outbreaks because they share sourcing, preparation, and distribution of foods. Public reporting of illness via formal channels is limited, whereas social media platforms host abundant user-generated content that can provide timely public health signals. This paper analyzes signals from Yelp reviews produced by a Hierarchical Sigmoid Attention Network (HSAN) classifier and compares them with official restaurant inspection outcomes issued by the New York City Department of Health and Mental Hygiene (NYC DOHMH) in 2023. We evaluate correlations at the Census tract level, compare distributions of HSAN scores by prevalence of C-graded restaurants, and map spatial patterns across NYC. We find minimal correlation between HSAN signals and inspection scores at the tract level and no significant differences by number of C-graded restaurants. We discuss implications and outline next steps toward address-level analyses.
\end{abstract}

\section{Introduction}
Foodborne illness imposes substantial morbidity and economic burden globally, accounting for millions of cases of gastrointestinal disease each year and resulting in significant public health costs. Traditional surveillance systems primarily rely on clinical reports, laboratory confirmations, and regulatory inspections. While these systems are essential for official monitoring and response, they often suffer from delays and underreporting because affected individuals do not always seek medical attention or submit formal complaints. Consequently, many potential outbreaks go undetected until they become widespread.

To address this limitation, social media data offer an alternative and complementary channel for detecting early signals of foodborne illness. User-generated platforms such as Yelp provide spontaneous, real-time accounts of dining experiences, some of which include descriptions of suspected illness after restaurant visits. The \textit{Adaptive Information Extraction from Social Media for Actionable Inferences in Public Health} project leverages these data to augment existing surveillance mechanisms. In this collaboration between Columbia University and the New York City Department of Health and Mental Hygiene (NYC DOHMH), Yelp reviews are retrieved daily and analyzed using a Hierarchical Sigmoid Attention Network (HSAN) model \cite{karamanolakis2019wnut}. The HSAN classifier assigns each review a probability score indicating its likelihood of describing a foodborne illness event. Reviews exceeding a predefined threshold are then transmitted to NYC DOHMH for further evaluation alongside traditional inspection workflows.

This pipeline enables the near real-time monitoring of potential public health threats and helps prioritize inspections based on emerging online patterns. The present study investigates whether HSAN-derived social media signals align with official NYC DOHMH restaurant inspection outcomes at the Census tract level for 2023. By quantifying these relationships spatially and statistically, we aim to assess whether user-reported illness mentions can serve as a meaningful proxy for actual food safety conditions, potentially strengthening hybrid surveillance models that integrate machine learning with regulatory oversight.

\section{Data and Methods}

\subsection{Data Sources}
This study integrated three primary datasets. Yelp restaurant reviews from establishments across the five boroughs of New York City were obtained through a private API partnership for the 2023 calendar year. Each record contained textual content, review date, user rating, restaurant metadata, and geographic coordinates. Official restaurant inspection results were sourced from the New York City Department of Health and Mental Hygiene (NYC DOHMH) Open Data portal, which provides both numeric inspection scores—where higher scores indicate poorer sanitary outcomes—and corresponding letter grades (A, B, or C). Census tract boundary shapefiles were downloaded from the U.S.\ Census Bureau's TIGER/Line dataset to ensure a consistent geographic framework for aggregation. Restaurant addresses were geocoded using the OpenStreetMap (OSM) Nominatim service, and duplicate or incomplete records were removed before spatial joining to maintain data accuracy.

\subsection{HSAN Scoring}
Each Yelp review was processed using a pre-trained \textit{Hierarchical Sigmoid Attention Network} (HSAN) classifier \cite{karamanolakis2019wnut}, a weakly supervised neural architecture designed to detect potential foodborne illness mentions in user-generated text. The HSAN model captures contextual information through a hierarchical attention mechanism, identifying relevant tokens and sentences while outputting a continuous probability score between 0 and 1 via a sigmoid activation layer. For each review $r_i$, the model produced a likelihood value $p_i$ reflecting the probability that the review described a foodborne illness event. Reviews with $p_i \ge 0.05$ were retained for analysis, a threshold chosen to balance false positives and recall based on prior validation studies. Model inference was performed using Python 3.10 and the PyTorch framework on Google Colab.

\subsection{Aggregation}
To enable spatial comparison between text-based illness signals and official inspection results, both datasets were aggregated at the Census tract level. Each geocoded restaurant was assigned to its corresponding tract polygon using a spatial join in GeoPandas. For every tract $t_j$, we computed the mean HSAN probability $\overline{p_j}$ across all qualifying reviews and the mean inspection score $\overline{s_j}$ across all inspected restaurants within the same tract. Tracts with fewer than three valid reviews or two inspections were excluded to reduce statistical instability arising from small sample sizes. All spatial processing, including coordinate transformations and boundary clipping, was performed using QGIS 3.34 and GeoPandas 0.14.

\subsection{Analyses}
Spatial visualization and statistical testing were carried out to explore the relationship between social media–derived indicators and inspection outcomes. Choropleth maps of tract-level mean HSAN and inspection scores were created, along with a kernel-density heatmap weighted by HSAN probabilities to visualize concentrations of illness-related reviews. A bivariate choropleth map was also generated to highlight tracts where both indicators were simultaneously elevated. Pearson ($r$) and Spearman ($\rho$) correlation coefficients were calculated between $\overline{p_j}$ and $\overline{s_j}$ to assess both linear and rank-based associations, with separate analyses performed for each borough to evaluate spatial heterogeneity. Additionally, the distribution of tract mean HSAN scores was compared across three categories of restaurant environments—tracts containing 1--2, 3--4, or five or more C-graded restaurants—using the Kruskal–Wallis nonparametric test to determine whether statistically significant differences existed among groups. All analyses were conducted in Python 3.10 using \texttt{pandas}, \texttt{scipy}, and \texttt{numpy} for computation and visualization, with cartographic design completed in QGIS.

This integrated workflow links machine learning–based social media signal extraction with regulatory food safety data, facilitating both quantitative assessment and visual exploration of potential associations between public online discourse and official inspection outcomes across New York City.

\section{Results}

\subsection{Spatial Patterns}
\begin{figure}[h]
  \centering
  \begin{subfigure}{.32\linewidth}
    \centering\includegraphics[width=\linewidth]{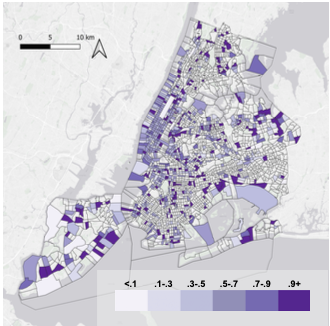}
    \caption{Mean HSAN score by tract.}
  \end{subfigure}\hfill
  \begin{subfigure}{.32\linewidth}
    \centering\includegraphics[width=\linewidth]{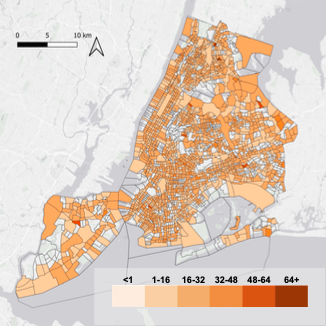}
    \caption{Mean inspection score by tract.}
  \end{subfigure}\hfill
  \begin{subfigure}{.32\linewidth}
    \centering\includegraphics[width=\linewidth]{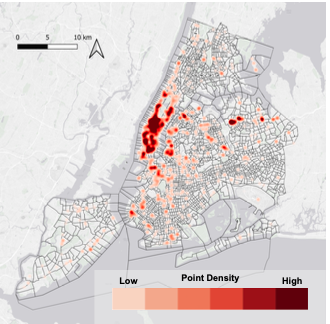}
    \caption{HSAN-weighted point density.}
  \end{subfigure}
  \caption{Univariate tract maps and review density.}
  \label{fig:maps}
\end{figure}

Figure~\ref{fig:maps} illustrates the spatial distribution of foodborne-illness–related Yelp reviews and official inspection outcomes across New York City. The bivariate map combines both indicators, while the univariate panels display each metric individually along with the HSAN-weighted point density of reviews. Brooklyn shows numerous tracts with relatively high HSAN and inspection scores, indicating areas with frequent illness mentions as well as less favorable inspection results. In contrast, Manhattan and Queens display greater heterogeneity, with clusters of high HSAN activity that do not consistently correspond to poor inspection outcomes. Staten Island and the Bronx exhibit fewer Yelp reviews overall, which likely reflects differences in population density and restaurant volume rather than genuine absence of risk.

These spatial patterns suggest that social media signals and inspection metrics vary considerably across boroughs, shaped by both demographic and infrastructural factors. The visualization highlights several local hotspots where both HSAN and inspection scores are elevated, implying potential alignment between online illness reports and observed sanitary deficiencies.

\begin{figure}[h]
  \centering
  \includegraphics[width=.75\linewidth]{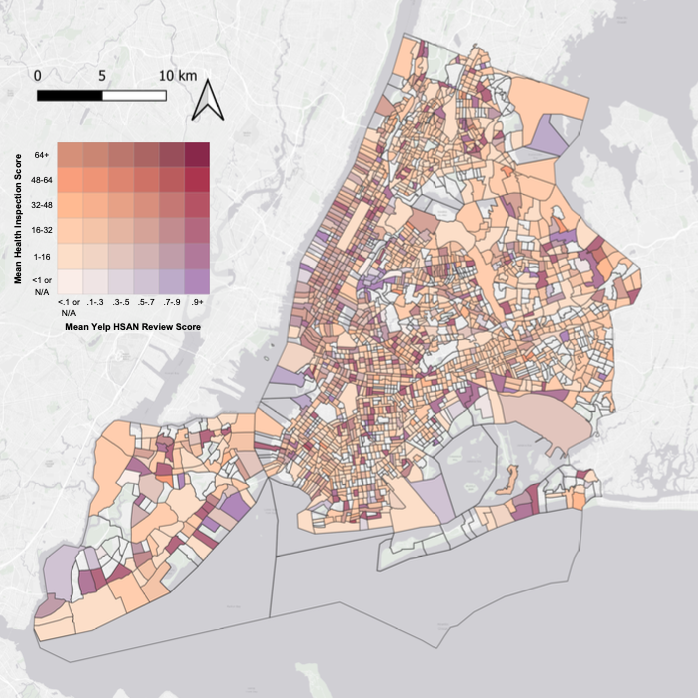}
  \caption{Bivariate choropleth of mean HSAN score (x-axis classes) and mean restaurant inspection score (y-axis classes) by NYC Census tract, 2023.}
  \label{fig:bivar}
\end{figure}

\subsection{Associations Between HSAN and Inspections}
\begin{figure}[h!]
  \centering
  \includegraphics[width=.78\linewidth]{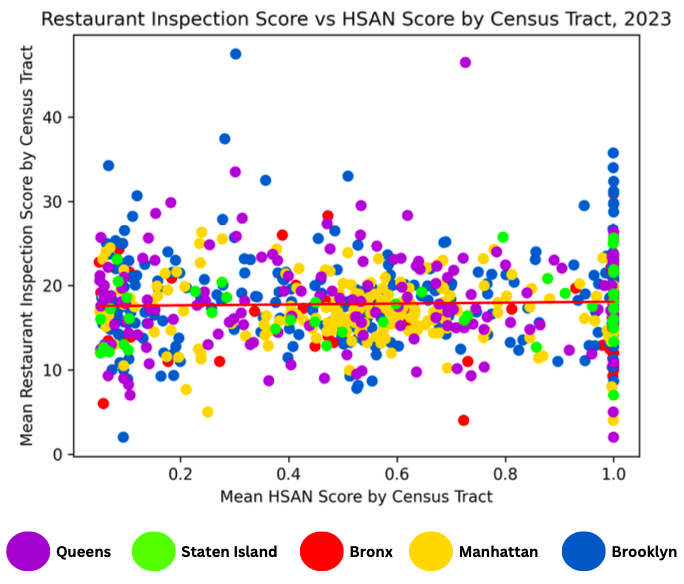}
  \caption{Scatter of tract mean HSAN score vs.\ mean inspection score by borough, 2023. Each point represents one Census tract.}
  \label{fig:scatter}
\end{figure}

Figure~\ref{fig:scatter} presents the relationship between tract-level mean HSAN probabilities and mean inspection scores. Each point represents a Census tract, color-coded by borough. The fitted regression line indicates a weak positive slope, suggesting only minimal linear association between social media illness mentions and inspection outcomes. Across the entire city, tracts with higher inspection scores (i.e., poorer health outcomes) do not necessarily correspond to tracts with high HSAN scores. Borough-specific patterns reveal subtle variations: Brooklyn and Staten Island show slightly stronger positive tendencies, while Manhattan exhibits a diffuse pattern with little correlation.

\begin{figure}[h]
  \centering
  \includegraphics[width=.7\linewidth]{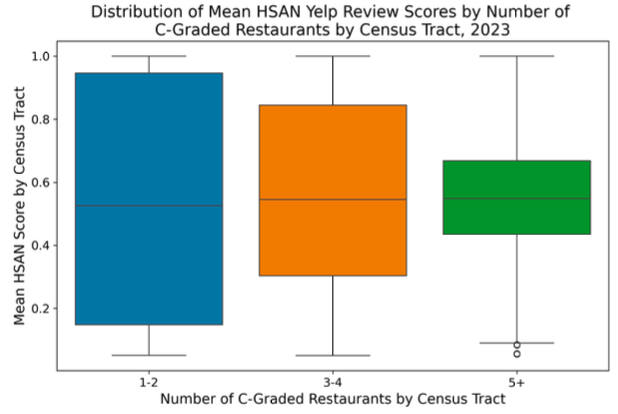}
  \caption{Distribution of tract mean HSAN scores by number of C-graded restaurants (1--2, 3--4, 5+).}
  \label{fig:box}
\end{figure}

The distribution of mean HSAN scores by number of C-graded restaurants (Figure~\ref{fig:box}) shows no statistically significant difference across groups, as confirmed by the Kruskal–Wallis test ($H \approx 1.4$, $p \approx 0.5$). Although tracts with five or more C-graded restaurants have slightly higher median HSAN values, the overlap of interquartile ranges indicates that Yelp-derived illness signals and official inspection grades are not directly aligned at this spatial scale.

\subsection{Correlation Coefficients}
Correlation analysis further supports the limited strength of association between social media signals and inspection performance. As summarized in Table~\ref{tab:corr}, both Pearson ($r$) and Spearman ($\rho$) correlation coefficients are small and statistically insignificant across the full dataset ($r = 0.03$, $p = 0.36$; $\rho = 0.05$, $p = 0.22$). Borough-level correlations follow a similar pattern: Staten Island exhibits the highest coefficients ($r = 0.22$, $\rho = 0.24$), while Manhattan and the Bronx show weak or negative associations. These findings suggest that, at the Census tract level, variation in HSAN-based illness likelihood is largely independent of mean inspection performance.

The absence of a strong correlation may reflect the inherently different natures of the two data sources. Yelp reviews capture subjective user experiences and may describe symptoms occurring days after a dining event or at a different location, whereas inspection scores assess objective sanitary conditions observed by officials at fixed intervals. Together, these results imply that while social media offers valuable qualitative insight into potential illness trends, its alignment with regulatory data may only emerge at finer spatial or temporal resolutions.

\begin{table}[h!]
\centering
\begin{tabular}{lcc}
\toprule
\textbf{Borough} & $\mathbf{r \ (p)}$ & $\boldsymbol{\rho \ (p)}$\\
\midrule
All (N=725)          & 0.03 (0.36) & 0.05 (0.22) \\
Queens (N=158)       & 0.02 (0.77) & 0.04 (0.63) \\
Staten Island (N=43) & 0.22 (0.16) & 0.24 (0.12) \\
Bronx (N=60)         & -0.05 (0.70) & -0.07 (0.59) \\
Manhattan (N=195)    & -0.07 (0.35) & -0.03 (0.72) \\
Brooklyn (N=269)     & 0.09 (0.14) & 0.10 (0.11) \\
\bottomrule
\end{tabular}
\caption{Pearson and Spearman correlations between tract mean HSAN and mean inspection scores (2023).}
\label{tab:corr}
\end{table}

Overall, these findings reveal that while both Yelp-derived and inspection-based indicators highlight localized areas of public health concern, their citywide relationship remains weak. The disparity suggests that social media surveillance captures complementary rather than redundant information compared to traditional inspection processes.

\section{Discussion}

\subsection{Interpretation of Findings}
At the Census tract level, we observed weak and statistically non-significant associations between Yelp-derived HSAN signals and official NYC DOHMH restaurant inspection outcomes. This lack of correlation suggests that social media–based indicators and regulatory inspections may capture complementary aspects of public health risk rather than identical constructs. HSAN identifies reviews where users self-report symptoms consistent with foodborne illness, reflecting perceived health experiences. In contrast, inspection scores represent formal evaluations of sanitary conditions at specific points in time. The two measures differ not only in their conceptual focus but also in their temporal and spatial alignment—symptoms may manifest days after dining, and users may not always attribute illness to the correct restaurant.  

Spatially, Brooklyn exhibited the highest counts of tracts with both elevated HSAN and inspection scores, suggesting potential clustering of areas with both poor inspection performance and frequent illness mentions. Manhattan showed substantial heterogeneity: while it had many high-HSAN tracts, these were not necessarily linked to poor inspection outcomes, likely reflecting the borough’s dense restaurant environment and higher online review activity. Staten Island and the Bronx, by contrast, displayed limited review volume, constraining the sensitivity of social media signal detection. These results collectively imply that the integration of social media data into food safety surveillance is promising but may require finer geographic resolution to reveal meaningful patterns.

\subsection{Limitations}
Several limitations should be considered when interpreting these findings. First, the HSAN classifier operates probabilistically, and the choice of a fixed threshold ($p_i \ge 0.05$) may introduce bias or attenuate sensitivity to true illness signals. Second, Yelp reviews are inherently subjective and demographically skewed toward more active users, introducing selection bias and limiting generalizability to the broader dining population. Third, geocoding accuracy and spatial aggregation to Census tracts may obscure address-level variation; the tract boundary may not always correspond to the restaurant catchment area. Additionally, inspection and review data are asynchronous—an inspection may occur weeks or months before or after a review describing illness. Such temporal misalignment could dilute potential associations between illness mentions and inspection outcomes. Finally, no adjustments were made for contextual variables such as restaurant type, cuisine, price level, or tract-level sociodemographic factors, all of which may influence both inspection scores and the propensity to post illness-related reviews.

\subsection{Future Work}
Future analyses will expand upon these results by examining relationships at finer spatial and temporal scales. Address-level analyses, in particular, will enable direct linkage between individual restaurants’ HSAN scores and their inspection outcomes, potentially uncovering correlations that are masked at the tract level. Temporal alignment of review timestamps and inspection dates will also allow exploration of lead–lag dynamics—specifically, whether elevated HSAN signals precede poor inspection outcomes, indicating early warning potential.  

We additionally plan to incorporate covariates such as restaurant density, cuisine category, and local demographic characteristics to control for confounding factors. Methodologically, ablation studies will be conducted to evaluate the sensitivity of results to different HSAN thresholds and to compare the classifier’s performance against alternative language models fine-tuned for foodborne illness detection. Integrating these approaches will help refine the use of social media as a complementary surveillance tool, supporting public health agencies in prioritizing inspections and identifying emerging risks more rapidly and equitably.

\section{Ethics and Data Use}
All data use complied with agreements with Yelp Inc. and NYC DOHMH. Only aggregate statistics are reported in this paper.

\section{Acknowledgements}
We thank collaborators at NYC DOHMH and data providers: Yelp Inc., U.S.\ Census Bureau, and NYC Open Data. This project was supported by the National Science Foundation under Grant IIS-1563785. The views expressed are those of the authors and do not necessarily reflect those of NSF.

\end{document}